\documentclass[fleqn,12pt]{article}
\usepackage{amsmath}
\usepackage{epsfig}
\usepackage{amsfonts,amssymb,latexsym}
\usepackage[vcentermath,enableskew]{youngtab}

\ifx\href\asklfhas\newcommand{\href}[2]{#2}\fi

\newdimen\squaresize \squaresize=12pt
\newdimen\thickness \thickness=0.7pt

\def\square#1{\hbox{\vrule width \thickness
   \vbox to \squaresize{\hrule height \thickness\vss
      \hbox to \squaresize{\hss#1\hss}
   \vss\hrule height\thickness}
\unskip\vrule width \thickness} \kern-\thickness}

\def\cut#1{\hbox{\vrule width-1 \thickness
   \vbox to \squaresize{\hrule height-1 \thickness\vss
      \hbox to \squaresize{\hss#1\hss}
   \vss\hrule height-1\thickness}
\unskip\vrule width +4 \thickness} \kern-\thickness}

\def\vsquare#1{\vbox{\square{$#1$}}\kern-\thickness}

\def\young#1{
\vbox{\smallskip\offinterlineskip \halign{&\vsquare{##}\cr #1}}}

\newcommand{\tinyyoung}[1]{
\squaresize=7pt \thickness=0.4pt \mbox{\tiny\young{#1}}
\squaresize=12pt \thickness=0.7pt}

\makeatletter
\newbox\slashbox \setbox\slashbox=\hbox{$/$}
\def\pFMslash#1{\setbox\@tempboxa=\hbox{$#1$}
  \@tempdima=0.5\wd\slashbox \advance\@tempdima 0.5\wd\@tempboxa
  \copy\slashbox \kern-\@tempdima \box\@tempboxa}

\makeatother

\newcommand{\ft}[2]{{\textstyle {\frac{#1}{#2}} }}

\newcommand{\be}{\begin{equation}}
\newcommand{\ee}{\end{equation}}
\newcommand{\ben}{\begin{displaymath}}
\newcommand{\een}{\end{displaymath}}
\newcommand{\ba}{\begin{eqnarray}}
\newcommand{\ea}{\end{eqnarray}}
\newcommand{\nn}{\nonumber}

\newcommand{\bean}{\begin{eqnarray*}}
\newcommand{\eean}{\end{eqnarray*}}

\newcommand{\mathon}{\mathversion{bold}}
\newcommand{\mathoff}{\mathversion{normal}}

\def\moth{\mathsurround=0pt}
\newdimen\zo \zo=0pt

\def\tick{\leaders\hrule height 0.5ex depth 0pt \hskip 0.5pt}
\def\upboxfill{$\moth \setbox\zo\hbox{\tick}%
  \hskip 2pt\hbox to 0pt{$\tick$\hss}\hrulefill \hbox to
6pt{$\tick$\hss}$}

\def\dtick{\leaders\hrule height .34pt depth .5ex \hskip 0.5pt}
\def\downboxfill{$\moth \setbox\zo\hbox{\dtick}%
  \hskip 2pt\hbox to 0pt{$\dtick$\hss}\hrulefill \hbox to
6pt{$\dtick$\hss}$}

\makeatletter \@addtoreset{equation}{section} \makeatother

\newcommand{\pls}{\!+\!}
\newcommand{\mis}{\!-\!}

\begin{document}

\thispagestyle{empty}

\begin{flushright}
CERN-TH/2004-234\\
DESY 04-229
\end{flushright}

\bigskip
\medskip

\begin{center}

\renewcommand{\thefootnote}{\fnsymbol{footnote}}

\mathon {\bf\Large
Higher spin holography for SYM in $d$ dimensions} \mathoff
\bigskip\bigskip

\addtocounter{footnote}{2} 
\textbf{
J.~F.~Morales\footnote{\texttt{Francisco.Morales.Morera@cern.ch}} and
H.~Samtleben\footnote{\texttt{Henning.Samtleben@desy.de}}}

\addtocounter{footnote}{-1} \vspace{.3cm} $^\fnsymbol{footnote}$ 
\textit{CERN, Theory Division\\
CH-1211, Geneva 23, Switzerland}

\addtocounter{footnote}{1} \vspace{.3cm} $^\fnsymbol{footnote}$ 
\textit{
II. Institut f\"ur Theoretische Physik, Universit\"at Hamburg\\
 Luruper Chaussee 149, D-22761 Hamburg, Germany}

\setcounter{footnote}{0}

\end{center}
\bigskip

\begin{abstract}
We derive the spectrum of gauge invariant operators for 
maximally supersymmetric Yang-Mills theories in $d$~dimensions.
After subtracting the tower of BPS multiplets,
states are shown to fall into
long multiplets of a hidden $SO(10,2)$ symmetry 
dressed by thirty-two supercharges. 
Their primaries organize into a
universal, i.e.~$d$-independent pattern. 
The results are in perfect agreement with those following from 
(naive) KK reduction of type II
strings on the warped AdS$\times S$ near-horizon 
geometry of Dp-branes. 
\end{abstract}

\vspace*{1.5cm}


\section{Introduction}

Holography between type II strings on AdS$_5\times S^5$
and ${\cal N}=4$ super Yang-Mills (SYM) theory
implies that there is a point in the string parameter space
at which the symmetry enhances to the infinite dimensional
higher spin algebra --- dual to SYM theory at vanishing 
coupling constant~$g_{\rm YM}=0\,$. 
In \cite{Bianchi:2003wx} a proposal for the string spectrum 
at this symmetry enhancement point was put forward.  
The results rely on the assumption that the
massive string spectrum organizes in $S^5$ Kaluza-Klein (KK)
towers on top of an $SO(6)$ gauged version
of the five-dimensional theory that follows from dimensional 
reduction of type II strings on a flat torus. 
This is similar to what happens for supergravity states 
which fall into an infinite tower of BPS 
multiplets~\cite{Gunaydin:1984fk} whose ground floor 
shares the physical degrees of freedom with ten-dimensional
type II supergravity.
The full string spectrum is then obtained by standard
KK techniques~\cite{Salam:1981xd}.
Masses of the string states were determined 
in~\cite{Beisert:2003te} by extrapolating 
pp-wave frequencies down to finite $J$ along the line 
of vanishing gauge coupling~$g_{\rm YM}=0$.
The resulting spectrum perfectly matches that of gauge invariant operators 
in the dual free ${\cal N}=4$  SYM theory, 
providing strong support in favor of these (naive) assumptions. 

On the gauge theory side,
gauge invariant SYM operators beyond the BPS bound generically
organize into long representations of the supersymmetry 
algebra. At $g_{\rm YM}=0$ some of these long multiplets are 
shortened~\cite{Dobrev:1985qz} 
and regroup into
infinite multiplets of the higher spin (HS) algebra, manifesting the HS
symmetry enhancement~\cite{Beisert:2004di} 
(see~\cite{Vasiliev:2004cm} 
for recent state of the art reports and complete lists of references on HS gauge theories). 
On the string side this corresponds to an infinite tower of string states coming 
down to zero mass.\footnote{Mass here is always understood in the sense 
of the AdS$_{5}$ background.}
Still, by far not all the string states become massless, as one would have concluded from a naive 
tensionless limit, but only those corresponding to the conserved currents which realize 
the HS algebra. This minimal HS gauge theory on AdS$_5$ realizes 
the $hs(2,2|4)$ higher 
spin algebra and was first studied in~\cite{Sezgin:2001yf}. 
HS currents are associated to doubletons
in SYM$_{4}$. The remaining states organize in either Goldstone multiplets, 
required for the Higgsing of the HS symmetry, 
or in genuinely massive multiplets of the superconformal algebra.
Reversely, when gauge interactions are turned on ($g_{\rm YM}\neq 0$) 
the HS symmetry
is broken. In the bulk
this corresponds to massless higher spin fields
acquiring mass via the pantagruelic Higgs mechanism (``grande bouffe''),
in which an infinite tower of Higgs 
particles is eaten by the infinite tower of massless multiplets in order
to become massive.

Independently 
of conformal symmetry a similar picture should be realized by any gauge theory 
which arises as the holographic counterpart of some bulk theory.
Here we consider  the simplest nonconformal scenario:  
strings on near horizon Dp-brane geometries. 
The boundary theories in these cases are maximally supersymmetric 
SYM$_{d}$ theories in $d=p+1$ dimensions~\cite{Itzhaki:1998dd} 
and for $p\not=3$, they are nonconformal at $g_{\rm YM}\neq 0$. 
On the string side this corresponds to the fact that
the Dp-brane near horizon geometry is not a direct but a warped product 
AdS$_{{d+1}}\times S^{9-d}\,$. 
Accordingly, the ground state in the $(p\!+\!2)$-dimensional 
effective bulk theory does not correspond to a pure AdS geometry
but rather to a domain wall solution. So far, only the supergravity
content of these so called domain wall/QFT 
correspondences~\cite{Boonstra:1998mp} 
has been explored. Supergravity in warped AdS$\times S$ spaces
can be studied with similar techniques like those used for pure  AdS$\times S$
geometries. The spectrum of chiral primaries~\cite{Morales:2002ys}
and certain two-point functions~\cite{Gherghetta:2001iv} have been determined 
and the results have been shown to agree with 
what is expected from holography. 

The aim of this letter is to extend the match between QFT/domain wall
spectra  beyond the supergravity level.
In particular, we apply the KK 
algorithm developed in~\cite{Bianchi:2003wx,Beisert:2003te} to 
strings moving on warped AdS$\times S$ near horizon Dp-brane geometries. 
Masses of the string states again follow from extrapolation of the pp-wave limits.
String theories on Dp-brane plane waves are solvable for any $p$ and
have been studied in 
detail in~\cite{Fuji:2002vs}. 
We show that the results are in perfect agreement with the spectrum of charges
and multiplicities of gauge invariant operators in $d$-dimensional SYM theory. 
On the gauge theory side, operators are counted via Polya theory 
following~\cite{Bianchi:2003wx} (see also~\cite{Sundborg:1999ue}).
Recent applications of Polya theory to the study of ${\cal N}=4$ SYM
can be found in~\cite{Aharony:2003sx}.

The letter is organized as follows. In section~2, we briefly review
the proposal of~\cite{Bianchi:2003wx,Beisert:2003te} for the
string spectrum on AdS$_5\times S^5$ at the higher
spin enhancement point and show how the proposal naturally extends
to the string spectra on (warped) AdS$_{d+1}\!\times\!S^{9-d}\,$
geometries. In section~3, we compute the 
spectrum of gauge invariant single-trace SYM operators in $d$ dimensions.
We show that after subtracting the tower of BPS multiplets, the SYM 
spectra take a universal form, manifestly covariant under an $SO(10,2)$ 
symmetry dressed with 32 supercharges. 
Signs of this underlying structure in string theory 
have appeared in various 
contexts~\cite{Gunaydin:1998bt} 
but it still remains to be fully elucidated.  
Comparing to the string spectra on the warped geometries
we find perfect agreement.
Two appendices contain the technical details about the partition functions
and the BPS multiplets.

\bigskip

\section{Strings on Dp-brane geometries}

In this section we consider the spectrum of KK descendants of massive string 
excitations on  warped AdS$_{d+1}\times S^{9-d}$ spaces.
It has been argued in~\cite{Bianchi:2003wx} that the string spectrum on AdS$_5\times S^5$
at the higher spin symmetry enhancement point may be put into the simple form
\ba {\cal Z}_{\rm string}&=&
{\cal Z}_{\rm sugra}+ 
{\cal T}_{\rm susy}\times{\cal T}_{\rm KK}\times{\cal Z}_{\rm flat}
\;,
\label{string5}
\ea
where ${\cal Z}_{\rm sugra}$ is the BPS tower comprising the massless string spectrum,
\ba
{\cal Z}_{\rm flat}&=& \sum_{\ell=1}^\infty\, ( {\rm vac}_\ell \times {\rm vac}_\ell) 
\;,
\nonumber
\ea
denotes the massive string spectrum in flat space after dividing out 
the $2^{16}$-dimensional long multiplet of the ten-dimensional type II
superalgebra, and
\ba
{\cal T}_{\rm susy} &=& \frac{(1-t^{1/2})^{16}}{(1-t)^{4}}
\;,
\qquad
{\cal T}_{\rm KK} ~=~ \frac{1-t^{2}}{(1-t)^{6}}
\;,
\nonumber
\ea
denote the fundamental long multiplet of $PSU(2,2|4)$ and the KK tower 
of completely symmetric $SO(6)$ vector representations, respectively. 
The variable $t$ here labels the quantum number $\Delta$
in $PSU(2,2|4)$ corresponding to the mass of the string states. 
The massive string spectrum thus manifestly organizes into long
multiplets while the massless part comes in the infinite
tower of BPS multiplets of $PSU(2,2|4)$.
The factors ${\cal T}_{\rm susy}$ and ${\cal T}_{\rm KK}$ 
organize the supersymmetry and KK descendants, respectively, and 
combine into the SO(10,2) covariant expression 
${\cal T}_{\rm SO(10,2)}\equiv{\cal T}_{\rm susy}\times{\cal T}_{\rm KK}\,$. 
In other words, spacetime
derivatives and KK descendants combine into ${\cal T}_{\rm SO(10,2)}$  to reconstruct the
ten-dimensional momentum. 
The factor ${\cal Z}_{\rm flat}$ in~(\ref{string5}) is obtained from the flat space string 
spectrum after an appropriate lift to $SO(10)\!\times\!SO(2)_{\Delta}$.
The extra quantum number $\Delta$ is obtained by 
the BMN inspired mass formula $\Delta-J=\nu$ with $J$ the light cone 
charge and $\nu$ the string occupation number  
(see~\cite{Beisert:2003te} for details and explicit expressions
of ${\cal Z}_{\rm flat}$ for the first string levels). 
Later on, $\Delta$ will be
related to the naive dimensions of gauge invariant operators
 in SYM$_d$. Together, this shows that 
the massive part of the string spectrum~(\ref{string5})
takes the manifestly $SO(10)\!\times\!SO(2)$ covariant form 
${\cal T}_{\rm SO(10,2)}\times{\cal Z}_{\rm flat}\,$.

The same line of argument now applies to the computation of
KK descendants of massive string states on $S^{9-d}\,$. The
natural proposal for the string spectrum on the 
warped AdS$_{d+1}\times S^{9-d}\,$
background thus is 
\ba 
{\cal Z}_{\rm string}^{(d)}&=&
{\cal Z}_{\rm sugra}^{(d)}~+~ 
{\cal T}^{(d)}_{\rm susy}\times {\cal T}^{(d)}_{\rm KK}\times
{\cal Z}_{\rm flat}~=~
{\cal Z}_{\rm sugra}^{(d)}~+~ 
{\cal T}_{\rm SO(10,2)}\times{\cal Z}_{\rm flat}\;.
\label{string}
\ea
Interestingly, of the full spectrum only the supergravity part 
is sensitive to the dimension of the sphere $S^{9-d}$ while for the massive part
of the spectrum different values of $d$ simply correspond to different 
decompositions of the universal multiplet ${\cal T}_{\rm SO(10,2)}$
into superderivatives and KK descendants.
Specifically, these towers are given by
\ba
{\cal T}_{\rm susy}^{(d)} &=&  
\frac{(1-t^{1/2})^{\bf 16_{c}}}{(1-t)^{\bf d}}\;,
\qquad
{\cal T}_{\rm KK}^{(d)} ~=~
\frac{(1-t^2)}{(1-t)^{\bf 10-d}}\;.
\label{kksusy}
\ea
Here and in the following we denote by ${\bf d}$ and ${\bf 10\!-\!d}$ the vector 
representations of $SO(d)$ and $SO(10\!-\!d)$, respectively. 
The completely symmetric/antisymmetric tensor products of these
representations are generated by the expansions
\ba
&&(1-t)^{\bf d}\equiv 1-{\bf d}\, t+({\bf d}\times {\bf d})_{\rm A} \, t^2+\ldots
\;,
\nn\\[1ex]
&&
\frac{1}{(1-t)^{\bf d}} \equiv 1+{\bf d}\, t+({\bf d}\times {\bf d})_{\rm S} \, t^2+\ldots
\;,
\label{exp}
\ea
and likewise for the ${\bf 10\!-\!d}\,$. The ${\bf 16_{c}}$ representation 
in~(\ref{kksusy}) is understood as branching of the corresponding
$SO(10)$ spinor representation under
$SO(10\!-\!d)\!\times\!SO(d)\,$.

In summary, the full spectrum of massive
string excitations on
AdS$_{d+1}\times S^{9-d}$ at the HS enhancement point 
can be written in the $SO(10)\!\times\!SO(2)$
covariant form~(\ref{string}) and is independent of the dimension $d\,$.
The dependence on $d$ exclusively arises from the supergravity or BPS part 
${\cal Z}_{\rm sugra}^{(d)}$ of the spectrum.  From the holographic
perspective this is a priori surprising. 
It implies that the total SYM$_{d}$ spectrum after subtracting 
its BPS part assumes a universal, $d$-independent form
organized in $SO(10)\!\times\!SO(2)$ representations.
In the next section we show that this is indeed the case.

\bigskip

\mathon
\section{SYM$_{d}$ cyclic words}
\mathoff

We now consider maximally supersymmetric $U(N)$ gauge theories in $d\le10$
dimensions defined by dimensional reduction of ${\cal N}\!=\!1$ SYM in
$d\!=\!10$.  The elementary fields are the gauge field $A_\mu$,
$(10\!-\!d)$ scalar fields $\phi_i$ and fermions 
$\psi_{\alpha,a}$, $\bar{\psi}_{\dot{\alpha},\dot{a}}$ all
in the adjoint representation of the $U(N)$ gauge group.
The gauge field and the scalars transform in the vector representations
of the Lorentz $SO(d-1,1)$ and ${\cal R}$-symmetry $SO(10-d)$ group,
respectively. The fermions are spinors of plus-plus or 
minus-minus chirality with respect to the two symmetry groups,
such that they combine into a single irreducible representation of the
ten-dimensional $SO(9,1)$.
To facilitate later comparison with the string spectrum we will 
consider the SYM theory on $\mathbb{R}\times S^{d-1}$ and organize
the spectrum according to the $SO(d)$ isometry group of the sphere.

Together with their derivatives and modulo their field equations 
the elementary fields can be encoded in the one-letter partition function 
${\cal Z}^{(d)}_1(t)$  to which we also refer as
the singleton representation: the set of ``letters''. 
The variable $t$ is counting the dimension of the SYM letters.
In general, gauge invariant operators are given in terms of
multi-trace combinations, i.e.~``sentences'' built from ``words''
(single-trace) made from an ``alphabet" of these letters.
Here we focus on the spectrum of single-trace operators, 
and therefore we count cyclic ``words".

The various contributions to ${\cal Z}^{(d)}_1(t)$ can be written as
\ba
{\cal D}^s \phi_{i}: &&
 t \, \sum_{s=0}^\infty \Big(\tinyyoung{1&2& & &s\cr}-{\rm traces}\Big)
 ~=~\frac{t\,(1-t^2)}{(1-t)^{\bf d}}\,({\bf 10\!-\!d})\;, 
 \nn\\[1ex]
{\cal D}^{s-1} {\cal F}:
&&  \sum_{s=1}^\infty \Big(\tinyyoung{1&2& & &s\cr \nu\cr}-{\rm traces}\Big)
~=~1-\frac{1-{\bf d}\, t\,(1-t^{2})-t^4}{(1-t)^{\bf d}}\;,
\nn\\[1ex]
{\cal D}^s \psi, {\cal D}^s \bar{\psi} :&&
\, \sum_{s=0}^\infty 
\Big(\tinyyoung{1&2& & &s\cr}\times \!\tinyyoung{\times\cr} -{\rm traces}\Big) 
~=~ -\,\frac{ {\bf 16}_{s}\,t^{3/2}-{\bf 16}_{c}\,t^{5/2}}{(1-t)^{\bf d}}
\;,
\label{z1c}
\ea
with the boxes $\tinyyoung{ \cr}$ and $\tinyyoung{\times\cr}$ representing the vector and
spinorial representation, respectively, of $SO(10\!-\!d)$, and 
denominators generate spacetime derivatives according to~(\ref{exp}).
More precisely,
the towers $1/(1\!-\!t)^{\bf d}$ account for $SO(d)$ descendants (derivatives
along $S^{d-1}$)
while $(1-t^2)$ removes the traces associated to ${\cal D}^2$ terms.
Finally, subtracting the ${\bf 16}_{c}\, t^{5/2}$ term imposes the Dirac 
equation ${\cal D}\psi=0$ on the fermionic modes.
Collecting all the terms from (\ref{z1c}), 
the singleton partition function in
$d$ dimensions takes the simple form
 \ba {\cal Z}_1^{(d)}(t) &=& 1-\frac{{\bf Z_S}(t)}{(1-t)^{\bf d}}
 \;,
 \label{z10}
 \ea
in terms of the $d$-independent characteristic function
\ba
{\bf Z_S}(t)\equiv {\bf 1}-{\bf 10}\, t +{\bf 16}_s\, t^{3/2} -{\bf 16}_c\,
t^{5/2}+{\bf 10}\, t^3-{\bf 1}\, t^4
\;.
\label{z1} \ea
The function ${\bf Z_S}(t)$ carries a natural $SO(10)\!\times\!SO(2)$ structure. This
hidden symmetry is broken in the singleton partition 
function~(\ref{z10}) only by explicit insertion of the $SO(d)$ descendants.
 
The spectrum of single-trace operators can then be determined by
counting the number of cyclic words via Polya's formula
\ba
{\cal Z}^{(d)}_{\rm SYM}(t)&=& -\sum_{m=1}^\infty
\frac{\varphi(m)}{m}\,\log \left[1-{\cal Z}^{(d)}_1(t^m)\right]
\;,
\label{polya}
\ea
with Euler's totient function $\varphi(m)\,$.
Plugging (\ref{z10}) into (\ref{polya}) one finds for the
 SYM partition function
\ba 
{\cal Z}^{(d)}_{\rm SYM}(t)
&=&
 -\sum_{m=1}^\infty \frac{\varphi(m)}{m}\,\log\left[{\bf Z_S}(t^m) \right]
+
\frac{t\,\partial_t\, (1-t)^{\bf d} }{(1-t)^{\bf d}}
\nonumber\\[1ex]
&=& {\cal Z}^{(0)}_{\rm SYM}(t)
+ \frac{t\,\partial_t\, (1-t)^{\bf d} }{(1-t)^{\bf d}}
\;.
\label{zsym} 
\ea
Notice that according to (\ref{exp}), the last term contains only a
finite number of completely antisymmetric forms and their $SO(d)$ descendants.
Remarkably, this is the only dependence of the total partition function on 
the space-time dimension~$d\,$
while 
\ba
{\cal Z}^{(0)}_{\rm SYM}(t)&=&
  -\sum_{m=1}^\infty \frac{\varphi(m)}{m}\,\log\left[{\bf Z_S}(t^m) \right]
  \;,
\label{z0sym}
\ea
comes with a manifest $SO(10)\!\times\!SO(2)$ structure.

According to our discussion above, holography
would require that the $d$-dependent part in (\ref{zsym}) 
originates exclusively from BPS states.
We will now show that this is indeed the case.   
The BPS multiplet in $d$ dimensions is constructed by acting on
the completely symmetric chiral primary $(n0000) \, t^n $ with 
the $8$ unbroken supercharges
and spacetime derivatives
\ba
{\cal Z}_{{\rm BPS}\,n}^{(d)}(t) &=& \frac{1}{(1-t)^{\bf d}}\,
 \sum_{\epsilon_s=0,1}\, {\rm dim }\left[ (n 0 0 0 0)+\epsilon_s 
Q^{\rm BPS}_s\right]\,
t^{n+{1/2}\sum_s \epsilon_s} \;,\label{BPS}\\[1ex]
&& \mbox{with}\quad
Q^{\rm BPS}_s\in\Big\{ (-\ft12,\pm\ft12, \pm\ft12,\pm\ft12,\pm\ft12)   \;
\Big|\; {\rm with~even~number~of~+} \Big\} \;.
\nn
\ea
Here $(w_1, w_2, w_3, w_4, w_5)$ denote the weights of the corresponding
$SO(10\!-\!d)\times SO(d)$ representation. 
The explicit results in the various dimensions (in the Dynkin basis) 
are given in tables I--V in
Appendix~\ref{ABPS} below.
It is interesting to note that,
omitting energies and $SO(2)\subset SO(10)$ charges along the light cone plane,
the result in any dimension can be compactly
written as $({\bf 8_v}-{\bf 8_s})({\bf 8_v}-{\bf 8_s})\times [n000]\,$
with the appropriate branching of the $SO(8)$ representations.
This was used in \cite{Morales:2002ys}
 to show that the BPS part of the SYM spectrum indeed
reproduces the supergravity spectrum ${\cal Z}_{\rm sugra}^{(d)}$ 
on AdS$_{d+1}\times S^{9-d}$ .

The total BPS spectrum is then obtained by summing over $n$
\footnote{Here, 
for convenience we include in the BPS tower 
the $n=1$ singleton multiplet associated 
to ${\cal N}=4$ SYM multiplet living on the AdS boundary.}
\ba
{\cal Z}_{\rm BPS}^{(d)}(t) &=& 
\sum_{n=1}^\infty\, {\cal Z}_{{\rm BPS}\,n}^{(d)}(t)\;.
\ea
The result can be resumed in the remarkably simple relation
among BPS towers in different spacetime dimensions:
\ba 
{\cal Z}_{\rm BPS}^{(d)}(t) &=&
{\cal Z}_{\rm BPS}^{(0)}(t)
 +\frac{t\,\partial_t\, (1-t)^{\bf d} }{(1-t)^{\bf d}}
\;.
\label{BPSd} 
\ea
This relation is checked using the explicit form of BPS towers in
Appendix~\ref{ABPS}. Combining (\ref{zsym}) and (\ref{BPSd}), 
the total SYM spectrum can be written as
\ba
{\cal Z}^{(d)}_{\rm SYM}(t)&=& 
{\cal Z}_{\rm BPS}^{(d)}(t)+{\cal Z}^{\rm SYM}_{\rm long}(t)
\;,
\label{sym}
\ea
with
\ba
{\cal Z}^{\rm SYM}_{\rm long}(t)
&\equiv&{\cal Z}_{\rm SYM}^{(0)}(t)-{\cal Z}_{\rm BPS}^{(0)}(t)
~=~
-\sum_{m=1}^\infty \frac{\varphi(m)}{m}\,\log\left[{\bf Z_S}(t^m) \right]
-{\cal Z}_{\rm BPS}^{(0)}(t)
\;,
\label{spectrum}
\ea
a $d$-independent and manifestly $SO(10)\!\times\!SO(2)$
covariant function describing the non-BPS part of the SYM spectrum.
This is exactly the form predicted by
holography from the string spectrum.
More precisely, comparing the expressions~(\ref{string}) and~(\ref{sym})
leads to the prediction
\ba
{\cal Z}^{\rm SYM}_{\rm long} &=&
{\cal T}_{\rm SO(10,2)}\times{\cal Z}_{\rm flat}
\;,
\label{pred}
\ea
directly in terms of $SO(10)\!\times\!SO(2)$ representations,
which no longer explicitly depends on the spacetime dimensions $d\,$.
The first few terms in the expansions of the functions appearing
in this relation are given in Appendix~\ref{parti}.
In~\cite{Beisert:2003te}  
the two sides of this equation were shown to agree for $d=4$
till $\Delta=10\,$. Since this relation is $d$-independent, the 
agreement for $d\neq 4$  finally is a direct consequence of the $d=4$ match.

We have thus shown that the spectrum of free SYM theory in $d$ dimensions
takes the form~(\ref{sym}) of a $d$-dependent tower of BPS multiplets
and a universal $d$-independent part that falls into $SO(10)\!\times\!SO(2)$ representations.
We have then used the results of~\cite{Bianchi:2003wx,Beisert:2003te} to show 
that this spectrum precisely agrees with
the spectrum of massive string excitations on the warped background
AdS$_{d+1}\times S^{9-d}\,$. 
The explicit match is given in Appendix~\ref{parti} till $\Delta=6$.

Let us finish by showing that the universal part ${\cal Z}^{\rm SYM}_{\rm long}$ 
of the SYM spectrum indeed shares the factor $(1-t^{1/2})^{\bf 16_{c}}$ with
the long multiplet ${\cal T}_{\rm SO(10,2)}\,$.
For this one needs to work with the full character polynomial 
\ba
(1-t^{1/2})^{\bf 16_{c}} &\equiv&
\prod_{\alpha=1}^{16}\,(1-{\bf y}^{\bf q_{\alpha}}\,t^{1/2})\;,
\qquad {\bf y}^{\bf q_{\alpha}}=\prod_{i}\,y_{i}^{q_{\alpha i}}\;,
\\[1ex]
&&
{\bf q_{\alpha}}\in
\Big\{ (\pm\ft12,\pm\ft12, \pm\ft12,\pm\ft12,\pm\ft12)  
\;\Big|\;  \; {\rm with~even~number~of~+} \Big\}
\;,
\nonumber
\ea
and show that ${\cal Z}_{\rm SYM}(t,{\bf y})$ has zeros at 
$t={\bf y}^{-2\bf q_{\alpha}}\,$ for any $\alpha=1,\ldots 16$. 
The following observation is then crucial:
at the particular values $t={\bf y}^{-2\bf q_{\alpha}}\,$ 
the character polynomial of the $SO(10)\!\times\!SO(2)$ 
function ${\bf Z_S}(t,{\bf y})$ from~(\ref{z1}) factorizes according to
\ba
{\bf Z_S}(t,{\bf y})\Big|_{t={\bf y}^{-2\bf q_{\alpha}}} &=&
\prod_{i=1}^{5}  \Big(1-y_{i}^{2q_{\alpha i}}\,{\bf y}^{-2\bf q_{\alpha}} \Big)
\;.
\ea
As a consequence, the infinite sum in (\ref{spectrum}) at these points can
 be explicitly evaluated as
 \ba
 {\cal Z}_{\rm SYM}^{(0)}(t,{\bf y})\Big|_{t={\bf y}^{-2\bf q_{\alpha}}} &\!=\!&
 -\sum_{m=1}^\infty 
 \frac{\varphi(m)}{m}\,\log\left[{\bf Z_S}(t^m,{\bf y}^{m}) 
 \right]\Big|_{t={\bf y}^{-2\bf q_{\alpha}}}
 =~
 \sum_{i=1}^{5}
 \frac{y_{i}^{2q_{\alpha i}}}
 {{\bf y}^{2\bf q_{\alpha}}-y_{i}^{2q_{\alpha i}}}
 \;.
 \nonumber
\ea
With the full character polynomial of the $d\!=\!0$ BPS tower 
from table~I, one then verifies that
\ba
 {\cal Z}^{\rm SYM}_{\rm long}(t\!=\!{\bf y}^{-2\bf q_{\alpha}},{\bf y}) &=&
 {\cal Z}_{\rm SYM}^{(0)}(t\!=\!{\bf y}^{-2\bf q_{\alpha}},{\bf y})-
 {\cal Z}_{\rm BPS}^{(0)}(t\!=\!{\bf y}^{-2\bf q_{\alpha}},{\bf y})~=~ 0\;,
\ea
i.e.~the non-BPS part ${\cal Z}^{\rm SYM}_{\rm long}$
of the SYM spectrum indeed organizes into long multiplets
composed of $2^{16}\times {\rm dim}({\sf hws})$ states and their 
$SO(10,2)$ descendants.  
Note that in the counting we have not marked the length $L$
of SYM words. Indeed, the long multiplets 
in general combine SYM words of 
length $L$, $L\!+\!1$, and $L\!+\!2\,$~\cite{Beisert:2004di}.
Accordingly, there is no trace of this ``quantum number" in the string side.
This is not surprising since the length $L$ is not a real quantum number after
interactions are turned on.

\paragraph{Acknowledgements}
This work is supported by the DFG  grant SA 1336/1-1, 
and the European RTN Program MRTN-CT-2004-503369. We thank 
L. Andrianopoli, M. Bianchi, S. Ferrara and M.A. Lled\'o for discussions.

\bigskip

\begin{appendix}

\section*{Appendix}

\section{Partition functions}
\label{parti}

Here we list the blind partition functions 
appearing in both sides of equation~(\ref{pred}).
The left hand side is defined in terms of the $d=0$ SYM spectrum
according to~(\ref{spectrum}).
The BPS$_{0}$ polynomial follows from table~I
using the SO(10) multiplicity formula~(\ref{so10}) and summing over all $n$.
The SYM$_{0}$ partition function is given by (\ref{z0sym}).
Together, one finds
\ba
Z_{\rm SYM}^{0} &=& 
10 \, t-16 \, {t}^{3/2}+55 \, {t}^{2}-\hbox{144} \, {t}^{5/2}+\hbox{450} \, {t}^{3}-
\hbox{1440} \, {t}^{7/2}+\hbox{4735} \, {t}^{4}-\hbox{15616} \, {t}^{9/2}
+\nonumber\\
&&{}+
\hbox{52354} \, {t}^{5}-\hbox{177840} \, {t}^{11/2}+\hbox{608655} \, {t}^{6}
- \dots
\nn\\[2ex]
Z_{\rm BPS}^{(0)}&=&  \frac{t \, \left ( 10+64 \, \sqrt{t}+\hbox{196} \, t
+\hbox{352} \, {t}^{3/2}+\hbox{406} \, {t}^{2}+\hbox{304} \, {t}^{5/2}
+\hbox{145} \, {t}^{3}+40 \, {t}^{7/2}+5 \, {t}^{4}\right ) }
{{\left ( 1- t\right ) } \, {\left ( 1+\sqrt{t}\right ) }^{9}}\;.
\nonumber\\
\label{lhs}
\ea
The right hand side of (\ref{pred}) counts KK and supersymmetric descendants 
of the type II string spectrum. 
The on-shell spectrum of the ten-dimensional string in flat space has
been written in a manifestly $SO(10)\times SO(2)$ covariant form in 
Appendix~B of~\cite{Beisert:2003te}.
Masses were derived by extrapolating BMN frequencies down to finite $J$. 
Evaluating the resulting partition function one finds    
\ba
\sum_{\ell} |{\rm vac}_\ell|^2 &=& 
t^2+ \left ( 10 \, t^2- t^3\right )^{2}+ 
\left ( -16 \, {t}^{5/2}+54 \, {t}^{3}
-10 \, {t}^{4}\right )^{2}+
\nn\\[-1ex]
&&  {}+ \left ( 45 \, {t}^{3}-\hbox{144} \, {t}^{7/2}+\hbox{210} \, {t}^{4}
+16 \, {t}^{9/2}-54 \, {t}^{5}\right )^{2}
+ \left(t^{3}+\ldots \right)^{2}
+\ldots \;,\nn\\[2ex]
{\cal T}_{\rm susy}^{(d)}\, {\cal T}_{\rm KK}^{(d)} &=&  
\frac{(1-t^{1/2})^{\bf 16_{c}} (1-t^2)}{(1-t)^{\bf 10}}
\;.
\label{vacell}
\ea
In (\ref{lhs}), (\ref{vacell}) we listed the expansions relevant
for comparisons $\Delta\leq 6$ (but the agreement was checked all the
way till $\Delta=10$ in~\cite{Beisert:2003te}).

\section{BPS multiplets}
\label{ABPS}
In this appendix we summarize the BPS multiplets in $d\!=\!2k$ even 
spacetime dimensions computed from~(\ref{BPS}) and
organized under the group $SO(10\!-\!d)\!\times\!SO(d)\times SO(2)_{\Delta}\,$.
{} From their explicit form it is straightforward 
to verify the relation~(\ref{BPSd}) among BPS towers in different dimensions.

\bigskip
\bigskip
 
\begin{minipage}{15cm}
\begin{tabular}{|c|l|} \hline
$\Delta$ &\\ \hline
$n$ &
\footnotesize $[n,0000]$
\\[.5ex]
     $n\pls\ft12$ &
\footnotesize  $[n\mis1,0001]$
\\[.5ex]
     $n\pls1$ &
\footnotesize $[n\mis2,0100]$
\\[.5ex]
     $n\pls\ft32$ &
\footnotesize $[n\mis3,1010]$
\\[.5ex]
     $n\pls2$ &
\footnotesize $[n\mis3,0020]+[n\mis4,2000]$
\\[.5ex]
     $n\pls\ft52$ &
\footnotesize $[n\mis4,1010]$
\\[.5ex]
     $n\pls3$ &
\footnotesize $[n\mis4,0100]$
\\[.5ex]
     $n\pls\ft72$ &
\footnotesize $[n\mis4,0001]$
\\[.5ex]
     $n\pls4$ &
\footnotesize $[n\mis4,0000]$
\\
 \hline
 \end{tabular}
\medskip
 
 Table I: $d=0$ BPS multiplet $[n,0000]$ under $SO(10)$. 
\bigskip

{\footnotesize
\begin{tabular}{|c|l|} \hline
$\Delta$ &\\ \hline
$n$ &
\footnotesize $[n,000](0)$
\\[.5ex]
     $n\pls\ft12$ &
\footnotesize  $[n\mis1,001](+1)+[n\mis1,010](-1)$
\\[.5ex]
     $n\pls1$ &
\footnotesize $[n\mis2,100](+2)+[n\mis1,000](0)+[n\mis2,011](0)+[n\mis2,100](-2)$
\\[.5ex]
     $n\pls\ft32$ &
\footnotesize $[n\mis2,010](+3)+[n\mis2,001](+1)+[n\mis3,110](+1)+
[n\mis2,010](-1)+[n\mis3,101](-1)$\\
& $+[n\mis2,001](-3)$
\\[.5ex]
     $n\pls2$ &
\footnotesize $[n\mis2,000](+4)+[n\mis3,100](+2)+[n\mis3,020](+2)+
[n\mis2,000](0)+[n\mis3,011](0)$\\
&
$+[n\mis4,200](0)+[n\mis3,100](-2)+[n\mis3,002](-2)+[n\mis2,000](-4)$
\\[.5ex]
     $n\pls\ft52$ &
\footnotesize $[n\mis3,010](+3)+[n\mis3,001](+1)+[n\mis4,110](+1)+
[n\mis3,010](-1)+[n\mis4,101](-1)$\\
& $+[n\mis3,001](-3)$
\\[.5ex]
     $n\pls3$ &
\footnotesize $[n\mis4,100](+2)+[n\mis3,000](0)+[n\mis4,011](0)+[n\mis4,100](-2)$
\\[.5ex]
     $n\pls\ft72$ &
\footnotesize $[n\mis4,001](+1)+[n\mis4,010](-1)$
\\[.5ex]
     $n\pls4$ &
\footnotesize $[n\mis4,000](0)$
\\
 \hline
\end{tabular}}
\medskip

Table II: $d=2$ BPS multiplet $[n,000](0)$ under $SO(8)\times SO(2)$.  
\bigskip
\end{minipage}

\begin{minipage}{15cm}

{\footnotesize
\begin{tabular}{|c|l|} \hline
     $\Delta$ &\\ \hline
     $n$ & $[n,00]{(0\,0)}$ \\[.5ex]
     $n\pls\ft12$ & $[n\mis1,10]{(0\,\frac12)}+
[n\mis1,01]{(\frac12\,0)}$ \\[.5ex]
     $n\pls1$ & $[n\mis2,02]{(0\,0)}+[n\mis2,20]{(0\,0)}
+[n\mis1,00]{(0\,1)}+[n\mis1,00]{(1\,0)}
+[n\mis2,11]{(\frac12\,\frac12)}$ \\[.5ex]
     $n\pls\ft32$ & $[n\mis2,10]{(0\,\frac12)}+[n\mis3,12]{(0\,\frac12)}
+[n\mis2,01]{(\frac12\,0)}+[n\mis3,21]{(\frac12\,0)}+
[n\mis2,01]{(\frac12\,1)}$\\
&{} $
+[n\mis2,10]{(1\,\frac12)}$ \\[.5ex]
     $n\pls2$ & $2\!\cdot\![n\mis2,00]{(0\,0)}+[n\mis4,22]{(0\,0)}
+[n\mis3,02]{(0\,1)}+[n\mis3,20]{(1\,0)}+2\!\cdot\![n\mis3,11]{(\frac12\,\frac12)}$\\
     & $+[n\mis2,00]{(1\,1)}$
\\[.5ex]
$n\pls\ft52$ &
$[n\mis3,10]{(0\,\frac12)}+[n\mis4,12]{(0\,\frac12)}+
[n\mis3,01]{(\frac12\,0)}+[n\mis4,21]{(\frac12\,0)}
+[n\mis3,01]{(\frac12\,1)}$\\
&{} $+[n\mis3,10]{(1\,\frac12)}$ \\[.5ex]
     $n\pls3$ & $[n\mis4,02]{(0\,0)}+[n\mis4,20]{(0\,0)}
+[n\mis3,00]{(0\,1)}+[n\mis3,00]{(1\,0)}
+[n\mis4,11]{(\frac12\,\frac12)}$ \\[.5ex]
     $n\pls\ft{7}2$ & $[n\mis4,10]{(0\,\frac12)}+
[n\mis4,01]{(\frac12\,0)}$ \\[.5ex]
     $n\pls4$ & $[n\mis4,00]{(0\,0)}$ \\
\hline
\end{tabular}
}
\medskip

Table III: $d=4$ BPS multiplet $[n,00](00)$ under $SO(6)\times SO(4)$.
\bigskip

{\footnotesize
\begin{tabular}{|c|l|} \hline
     $\Delta$ &\\ \hline
     $n$ & $[n,n]{(000)}$ \\[.5ex]
     $n\pls\ft12$ & $[n,n\mis1]{(001)}+
[n\mis1,n]{(010)}$ \\[.5ex]
     $n\pls1$ & $[n,n\mis2]{(100)}+[n\mis2,n]{(100)}
+[n\mis1,n\mis1]{(011)}+[n\mis1,n\mis1]{(000)}$ \\[.5ex]
     $n\pls\ft32$ &
  $[n\mis2,n\mis1]{(101)}+[n\mis1,n\mis2]{(110)}+[n\mis1,n\mis2]{(001)}+[n\mis2,n\mis1]{(010)}$\\
&{} $+[n,n\mis3]{(010)}+[n\mis3,n]{(001)}$ \\[.5ex]
     $n\pls2$ &
     $[n\mis2,n\mis2](000)+[n\mis2,n\mis2](200)+[n\mis2,n\mis2](011)
     +[n\mis3,n\mis1](002)$\\
&{} $+[n\mis3,n\mis1](100)
+[n\mis1,n\mis3](020)+[n\mis1,n\mis3](100)+[n,n\mis4](000)+[n\mis4,n](000)$
\\[.5ex]
$n\pls\ft52$ &
  $[n\mis3,n\mis2]{(101)}+[n\mis2,n\mis3]{(110)}+[n\mis2,n\mis3]{(001)}+[n\mis3,n\mis2]{(010)}$\\
&{} $
  +[n\mis1,n\mis4]{(010)}+[n\mis4,n\mis1]{(001)}$ \\[.5ex]
     $n\pls3$ & $[n\mis2,n\mis4]{(100)}+[n\mis4,n\mis2]{(100)}
+[n\mis3,n\mis3]{(000)}+[n\mis3,n\mis3]{(011)}$  \\[.5ex]
     $n\pls\ft{7}2$ & $[n\mis3,n\mis4]{(001)}+
[n\mis4,n\mis3]{(010)}$ \\[.5ex]
     $n\pls4$ & $[n\mis4,n\mis4]{(000)}$ \\
\hline
\end{tabular}
}

Table IV: $d=6$ BPS multiplet $[n,n](000)$ under $SO(4)\times SO(6)$.
\bigskip

\begin{tabular}{|c|l|} \hline
$\Delta$ &\\ \hline
$n$ &
\footnotesize $[+2n](0000)+[-2n](0000)$
\\[.5ex]
     $n\pls\ft12$ &
\footnotesize  $[+2n\mis1](0001)+[-2n\pls1](0010)$
\\[.5ex]
     $n\pls1$ &
\footnotesize $[+2n\mis2](0100)+[-2n\pls2](0100)$
\\[.5ex]
     $n\pls\ft32$ &
\footnotesize $[+2n\mis3](1010)+[-2n\pls3](1001)$
\\[.5ex]
     $n\pls2$ &
\footnotesize  $[+2n\mis4](0020)+[-2n\pls4](0002)+[+2n\mis4](2000)+[-2n\pls4](2000)$
\\[.5ex]
     $n\pls\ft52$ &
\footnotesize  $[+2n\mis5](1010)+[-2n\pls5](1001)$
\\[.5ex]
     $n\pls3$ &
\footnotesize $[+2n\mis6](0100)+[-2n\pls6](0100)$
\\[.5ex]
     $n\pls\ft72$ &
\footnotesize $[+2n\mis7](0001)+[-2n\pls7](0010)$
\\[.5ex]
     $n\pls4$ &
\footnotesize $[+2n\mis8](0000)+[-2n\pls8](0000)$
\\
 \hline
\end{tabular}
\medskip

Table V: $d=8$ BPS multiplet $[+2n](0000)$ under $SO(2)\times SO(8)$. 

\end{minipage}
\bigskip
\bigskip

The dimensions of $SO(10\!-\!d)\!\times\!SO(d)$ representations
follow (upon reduction) from
the $SO(10)$ multiplicity formula \footnote{More precisely
$SO(2m)$ multiplicities are given by  suppressing any term in the product
(\ref{so10}) involving $n_{i}$ with $i>m$ and the corresponding factorials.} 
\ba
\lefteqn {{\rm dim}[n_1,n_2,n_3,n_4,n_5]~=~\frac{1}{7! \, 5! \, 3! \, 4!}\, 
\left ( 1\!+\!n1\right )   \left ( 1\!+\!n2\right )   \left ( 1\!+\!n3\right )   \left ( 1\!+\!n4\right )   
\left ( 1\!+\!n5\right )  \times }\label{so10}\\
&&\times \left ( 2\!+\!n2\!+\!n3\right )   \left ( 2\!+\!n1\!+\!n2\right )   
\left ( 2\!+\!n3\!+\!n4\right )   \left ( 2\!+\!n3\!+\!n5\right )   
\left ( 3\!+\!n1\!+\!n2\!+\!n3\right )  \times
\nn\\
&&\times \left ( 3\!+\!n2\!+\!n3\!+\!n4\right )   \left ( 3\!+\!n2\!+\!n3\!+\!n5\right )   
\left ( 3\!+\!n3\!+\!n4\!+\!n5\right )   \left ( 4\!+\!n1\!+\!n2\!+\!n3\!+\!n4\right ) 
\times 
\nn\\
&&\times \left ( 4\!+\!n1\!+\!n2\!+\!n3\!+\!n5\right )  
 \left ( 4\!+\!n2\!+\!n3\!+\!n4\!+\!n5\right )    
 \left ( 5\!+\!n1\!+\!n2\!+\!n3\!+\!n4\!+\!n5\right )  \times
 \nn\\&&\times \left ( 5\!+\!n2\!+\!2  n3\!+\!n4\!+\!n5\right )  
\left ( 6\!+\!n1\!+\!n2\!+\!2  n3\!+\!n4\!+\!n5\right )  
 \left ( 7\!+\!n1\!+\!2  n2\!+\!2  n3\!+\!n4\!+\!n5\right )\;.
 \nn
\ea

\end{appendix}


\providecommand{\href}[2]{#2}\begingroup\raggedright\endgroup

\end{document}